\documentclass{article}
\usepackage[english]{babel}
\usepackage[utf8]{inputenc}
\usepackage{amsmath}
\usepackage{graphicx}
\usepackage[colorinlistoftodos]{todonotes}
\usepackage{sidecap}
\usepackage{subcaption}
\usepackage{epstopdf}
\usepackage{authblk}

 \usepackage{lmodern} 
 \usepackage[T1]{fontenc}
 \usepackage{amssymb}
 \numberwithin{equation}{section}
 \usepackage{longtable}
 \usepackage{enumerate}
 \usepackage{cite}
 \usepackage{tikz}
 \usetikzlibrary{shapes,arrows}
\begin{document}
\title{Particle-fluid-structure interaction for debris flow impact on flexible barriers}
\author{Alessandro Leonardi}
\author{Falk K. Wittel}
\author{Miller Mendoza}
\author{Roman Vetter}
\author{Hans J. Herrmann}
\affil{Institute for Building Materials, ETH Zurich\\
Computational Physics for Engineering Materials\\
        Stefano-Franscini-Platz 3, CH-8093 Zurich, Switzerland\\
        \textit{aleonardi@ethz.ch}}
\maketitle

\begin{abstract}
Flexible barriers are increasingly used for the protection from debris flow in mountainous terrain due to their low cost and environmental impact. However, a numerical tool for rational design of such structures is still missing. In this work, a hybrid computational framework is presented, using a total Lagrangian formulation of the Finite Element Method (FEM) to represent a flexible barrier. The actions exerted on the structure by a debris flow are obtained from simultaneous simulations of the flow of a fluid-grain mixture, using two conveniently coupled solvers: the Discrete Element Method (DEM) governs the motion of the grains, while the free-surface non-Newtonian fluid phase is solved using the Lattice-Boltzmann Method (LBM). Simulations on realistic geometries show the dependence of the momentum transfer on the barrier on the composition of the debris flow, challenging typical assumptions made during the design process today. In particular, we demonstrate that both grains and fluid contribute in a non-negligible way to the momentum transfer. Moreover, we show how the flexibility of the barrier reduces its vulnerability to structural collapse, and how the stress is distributed on its fabric, highlighting potential weak points.
\end{abstract}

\section{Introduction}
\label{Intro}
Debris flows are among the most hazardous natural events, among other reasons due to their destructive potential, their unpredictability, and the difficulties in designing effective countermeasures \cite{Iverson1997,Hungr2005}. The development of hazard maps has contributed to reducing the risk in many mountainous areas, forbidding or limiting constructions in potentially dangerous areas \cite{VanWesten2005}. Often, these regulatory measures fail to reduce the risk to an acceptable level, especially for settlements already located in hazardous terrain or in situations where an unexpected event suddenly highlights a potential risk \cite{Wendeler}. In these cases, structural countermeasures, like barriers, piles, check dams \cite{Armanini1991} and detention basins are commonly employed to further reduce the risk. A rather new countermeasure are flexible barriers that are efficiently used for quick interventions  and for small basins. Structurally similar to the barriers used for snow avalanches and rockfall prevention \cite{Volkwein2009}, they often consist of one or more steel cable nets, spanning the whole width of the riverbed while being anchored to the channel banks. The advantages of flexible barriers over rigid ones is a drastic reduction in construction costs, as well as a a smaller environmental impact. They make optimal use of material and land and are easier to dismantle and substitute \cite{Wendeler2008a}.

For an efficient design, the impact pressure excerted from the debris flow on the barrier is of utter importance. However, acquiring reliable estimates is a challenging design issue. The hydrodynamic force transmitted by the flow to the structure mainly depends on the volume of the debris material, on the composition of the sediments and on the impact speed of the flowing mass \cite{Armanini1990,Bugnion2011a,Canelli2012,Brighenti2013}. While this is true for any retention measure, flexible cable nets have the additional complication of being  permeable to the fluid phase and to sediments smaller than the mesh size \cite{Geographie}. The presence of a grain-size distribution in the flowing sediments, however, implies that grains larger than the mesh spacing are impounded by the net, which in turn reduces permeability to smaller size portions. When the barrier is completely clogged, even the fluid can be prevented from passing through. Due to this feedback mechanism, the peak pressure can be shifted beyond the initial impact, and the dynamic load gets distributed over a longer time period. Moreover, the flexibility of the barrier provides another means of reducing the impact momentum, since the structure adapts to the received impulse. Therefore, the assumption at the base of protection structure design, namely the possibility of uncoupling the fluid and the structural problem, renders incorrect.

\begin{figure}[tb]
\centering
\includegraphics[]{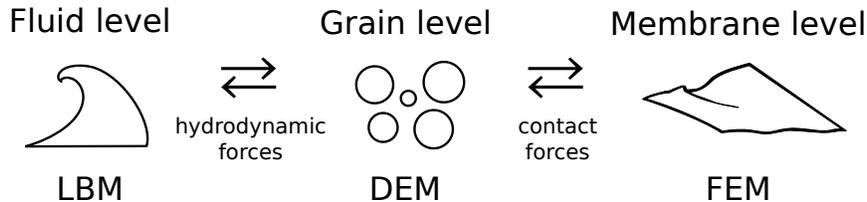}
\caption{\label{conceptual} Outline of the coupling scheme.}
\end{figure}

Flexible structures are relatively new compared to rigid ones and therefore lack a comprehensive set of experimental data to aid the rational design. Experiments have been carried out in Ref.~\cite{Roth2006}. Full-size experiments such as this are extremely expensive and therefore mainly single realizations. Downscaled experiments unfortunately face the problem of scaling the various physical phenomena involved (see Ref.~\cite{Scheidl2012} and references therein). The motivation for this work lies in providing a numerical approach, complete and at the same time efficient, to be used for the optimization of barrier design. We represent debris flow as a mixture of a granular and a non-Newtonian fluid phase. The grain dynamics is solved with the fluid by the Lattice-Boltzmann Method (LBM) (Sec.~\ref{LBM}) and the Discrete Element Method (DEM) (Sec.~\ref{DEM}). An outline of the coupling algorithm is given in Sec.~\ref{LBM-DEM}. The flexible barrier is modeled using the Finite Element Method (FEM) (Sec.~\ref{DEM-FEM}). It only interacts with the grains and not with the fluid. This reduces the computational cost while at the same time it mimics the real filter features of a cable net. When grains collide with the barrier, they mediate between the fluid and the barrier, hence transmitting the hydrodynamic force, see Fig.~\ref{conceptual}. Some snapshot from a simulation of this type are shown in Fig.~\ref{show} (c-e).

\section{Numerical Methods}

\subsection{Debris flow continuum phase: the LBM} \label{LBM}
\begin{figure}[t]
\centering
\includegraphics[width=\textwidth]{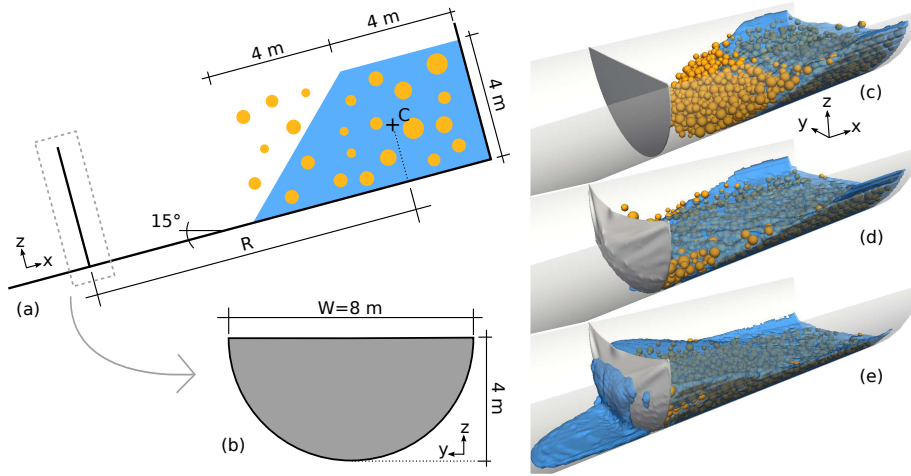}
\caption{\label{show} The setup of our simulation. (a) Geometry of the configuration before the simulation starts. Point $C$ is the debris center of mass. (b) Undeformed shape of the barrier. The snapshots on the right show the geometry of the flow before (c), during (d) and after (e) the impact on the barrier. The yellow spheres are the grains, the fluid free surface is depicted in blue, and the flexible barrier in grey.}
\end{figure}

The simulation of debris flow is commonly performed by the use of a continuum fluid approach, usually by using a non-Newtonian rheology to include the effect of the grains \cite{Iverson,Hutter1996}. Alternatively, debris flow is represented by a collection of grains, tracking the individual motion of each grain \cite{Teufelsbauer2009}. Recently, efforts have been devoted to trying to integrate these two approaches by describing the debris materials as a mixture of fluid and grains \cite{Sun2013,Leonardi2013}. The resulting numerical tool requires a continuum and a discrete solver. While still not frequently employed in geophysics, this approach was successfully applied for the simulation of other similar complex fluids, such as food or concrete \cite{Svec2012}. The method we use is similar to the one described in Ref.~\cite{Leonardi2014} and is therefore only briefly outlined here, starting from the fluid solver.

The LBM is a relatively recent approach to fluid dynamics where in contrast to Finite Volume Methods, conservation laws are not enforced on a continuum velocity and pressure field. Instead, the flowing mass is discretized as a collection of small colliding particles, represented by a probability distribution function $f$, which mimics the actual behavior of a fluid. The solution is made possible by a drastic reduction in the number of degrees of freedom, since the particles are only allowed to move on a fixed, regular grid, and with a velocity chosen among a discrete set $\{\boldsymbol{c}_i\}$. However, since mass and momentum conservation are nevertheless imposed, the outcome of an LBM simulation can be proven to be equivalent to a solution of the Navier-Stokes equations. The fluid mass density $\rho_\text{f}$, pressure $p_\text{f}$ and velocity $\boldsymbol{u}_\text{f}$ can be reconstructed starting from the discretized form of the distribution function $f_i$, as
 \begin{equation}
 \rho_\text{f}=\sum\limits_i f_i,\qquad
 p_\text{f}=c_s^2 \rho_\text{f},\qquad
 \boldsymbol{u}_\text{f}= \frac{1}{\rho_\text{f}}\sum\limits_i f_i\boldsymbol{c}_i,
 \label{densityVelocity}
\end{equation}
where $c_s$ is the speed of sound of the lattice and the sums run over all lattice sites $i$. The dynamics of the system is governed by the Lattice-Boltzmann equation which, assuming a temporal and spatial discretization  with unit spacing, reads
\begin{equation}
f_i (\boldsymbol{x}+ \boldsymbol{c}_i,t+1)= f_i (\boldsymbol{x},t)+\Omega_i (\boldsymbol{x},t) + F_i (\boldsymbol{x},t,\boldsymbol{F}) ,
\label{newfunctions}
\end{equation}
where $\Omega_i$ is the operator reconstructing the effect of molecular collisions. We express it using the Bhatnagar-Gross-Krook linear approximation \cite{Bhatnagar1954}, which drags the system towards the thermodynamic equilibrium state $f^{\text{eq}}_i$,
\begin{equation}
\Omega_i=\frac{f^{\text{eq}}_i-f_i}{\tau}.
\label{collision}
\end{equation}
The relaxation time $\tau$ governs the viscous behavior, being related to the viscosity of the fluid $\mu$ through
\begin{equation}
\tau=\frac{1}{2}+\frac{\mu}{c_s^2}.
\label{mu}
\end{equation}
To implement a non-Newtonian fluid, we set $\mu$ (and therefore $\tau$) to be a function of the shear rate, according to the approach in Refs.~\cite{Boyd2006,Svec2012}. The operator $F_i (\boldsymbol{x},t,\boldsymbol{F})$ in Eq.~\ref{newfunctions} implements the effects of external forcing terms $\boldsymbol{F}$ (see Sec.~\ref{LBM-DEM} for details).

\subsection{Debris flow granular phase: the DEM} \label{DEM}
The DEM is integrated with the LBM to obtain a hybrid representation of a debris flow, both as a continuum and as a discrete medium. Only a portion of the grain is represented, assuming spherical particles for simplicity. The motion of every grain is calculated by solving Newton's equations of motion for translational and rotational degrees of freedom in a fashion that has become standard for DEM simulations \cite{Poschel2005,Bicanic2007}. Whenever two grains come into contact, a repulsive force $\boldsymbol{F}_\mathrm{coll}$ is applied as a function of the overlap between the two grains
\begin{equation}
 \xi=r_1+r_2-\left\|\boldsymbol{d}_{1,2}\right\|,
\label{contact}
\end{equation}
where $\boldsymbol{d}_{1,2}$ denotes the distance vector between the center points of the grains, and $r_1$, $r_2$ their respective radii. The normal repulsive forces are calculated according to the Hertzian theory of viscoelastic collisions as
\begin{equation}
 F^\text{n}_{\text{coll}}=\frac{2}{3}\frac{E_\text{g}\sqrt{r_{\text{eff}}}}{\left(1-\nu_\text{g}^2\right)}
 \left(\xi^{3/2}+A\sqrt{\xi}\frac{d\xi}{dt}\right),
\label{hertz}
\end{equation}
with the Young's modulus $E_\text{g}$ and Poisson's ratio $\nu_\text{g}$ of the grains, while $A$ represents the damping constant  \cite{Brilliantov1996}. $r_{\text{eff}}$ is the effective radius defined as $r_{\text{eff}}=r_1 r_2 / (r_1+r_2)$. Two grains in contact exchange also tangential forces, proportional to their relative tangential velocity $u_{\text{rel}}^\text{t}$ and limited by Coulomb's friction law as
\begin{equation}
 F^\text{t}_{\text{coll}}=-\textrm{sign}\left(u_{\text{rel}}^\text{t}\right)\,
 \min\left\{\gamma\lvert u_{\text{rel}}^\text{t}\lvert, \tan (\psi) F^\text{n}_{\text{coll}}\right\},
\label{tang}
\end{equation}
where  $\psi$ is the dynamic friction angle and $\gamma$ is the shear damping coefficient.
Analogous principles are used for the solution of contacts with rigid walls or flexible obstacles. The forces arising from the collisions are added to the hydrodynamic interaction coming from the LBM, and to the gravitational force. The dynamics of the system is finally solved using a Gear predictor-corrector scheme \cite{Gear1971}.

\subsection{Thin shell with large deformations: the subdivision-surfaces FEM} \label{FEM}
The flexible barrier is represented by a FEM discretization of a thin shell in a total Lagrangian formulation. The underlying constitutive model follows the Kirchhoff-Love theory of thin shells, which is valid for shell thicknesses $h_\text{s}$ much smaller than the in-plane shell dimensions.
The method is only briefly outlined here and for further details we refer to Ref.~\cite{Vetter2013} and references therein. The middle surface of the shell is parametrized in both its stress-free reference ($\overline{\Omega}$) and deformed ($\Omega$) configuration. The indices $i,j,k,l=1,2$, denote covariant (subscripts) or contravariant (superscripts) components of vectors and tensors. Let  $\left\{ \theta^1,\theta^2,\theta^3\right\}$ be a curvilinear coordinate system. Any point on the middle surface can then written as $\overline{\boldsymbol{x}}\left(\theta^1,\theta^2\right)\in\overline{\Omega}$  when referring to the undeformed surface, and as $\boldsymbol{x}\left(\theta^1,\theta^2\right)\in\Omega$ when referring to the deformed surface. The position of arbitrary material points within the shell follows the same principles and is written as
\begin{equation}
\begin{aligned}
 \overline{\boldsymbol{p}}\left(\theta^1,\theta^2,\theta^3\right)&=\overline{\boldsymbol{x}}\left(\theta^1,\theta^2\right)+\theta^3
 \overline{\boldsymbol{a}}_3\left(\theta^1,\theta^2\right), \\
 {\boldsymbol{p}}\left(\theta^1,\theta^2,\theta^3\right)&={\boldsymbol{x}}\left(\theta^1,\theta^2\right)+\theta^3
 {\boldsymbol{a}}_3\left(\theta^1,\theta^2\right).
 \end{aligned}
\end{equation}
The terms $\overline{\boldsymbol{a}}_i$ and $\boldsymbol{a}_i$ are local directors for the surfaces $\overline{\Omega}$  and  $\Omega$. The first two components are tangent to the surface, and are obtained through differentiation as
\begin{equation}
 \overline{\boldsymbol{a}}_i\left(\theta^1,\theta^2\right)=\frac{\partial \overline{\boldsymbol{x}}}{\partial \theta^i},\qquad
  \boldsymbol{a}_i\left(\theta^1,\theta^2\right)=\frac{\partial \boldsymbol{x}}{\partial \theta^i}.
\end{equation}
The third component is computed according to the Kirchhoff hypotheses that straight material lines that are normal to the middle surface retain their straightness, normality and length in any deformed configuration, thus:
\begin{equation}
\overline{\boldsymbol{a}}_3=\frac{\overline{\boldsymbol{a}}_1 \times \overline{\boldsymbol{a}}_2}{\left\| \overline{\boldsymbol{a}}_1 \times \overline{\boldsymbol{a}}_2 \right\|},\qquad
\boldsymbol{a}_3=\frac{\boldsymbol{a}_1 \times \boldsymbol{a}_2}{\left\| \boldsymbol{a}_1 \times \boldsymbol{a}_2 \right\|}.
\end{equation}
With this formulation, strain measures can be expressed in a convenient form. We can obtain the covariant components of the first fundamental form as
\begin{equation}
\overline{a}_{i j} =\overline{\boldsymbol{a}}_i \cdot \overline{\boldsymbol{a}}_j,\qquad
a_{i j} =\boldsymbol{a}_i \cdot \boldsymbol{a}_j,
\end{equation}
and those of the second fundamental form as
\begin{equation}
\overline{b}_{i j} =\overline{\boldsymbol{a}}_3 \cdot \frac{\overline{\boldsymbol{a}}_i}{\theta^j},\qquad
b_{i j} =\boldsymbol{a}_3 \cdot \frac{\boldsymbol{a}_i}{\theta^j}.
\end{equation}
The in-plane ($2\times 2$) membrane and bending strain tensors in curvilinear coordinates then derive from the fundamental forms as 
\begin{equation}
\alpha_{ij}=\frac{1}{2}\left(a_{ij}-\overline{a}_{ij}\right),\qquad
\beta_{ij}=\overline{b}_{ij}-b_{ij}.
\end{equation}
With the further assumption for the shell material of being linearly elastic and therefore characterized only by a Young's modulus $E_\text{s}$ and Poisson's ratio $\nu_\text{s}$, the shell's total elastic energy is given by the integral of the Koiter energy density functional over the middle surface \cite{Koiter2008}
\begin{equation}
U_e \left[\overline{\boldsymbol{x}},\boldsymbol{x}\right]=
\frac{1}{2}\int_{\overline{\Omega}} 
h_\text{s} \alpha_{i j} C^{i j k l} \alpha_{k l}
+\frac{h_\text{s}^3}{12}\beta_{i j} C^{i j k l} \beta_{k l}
 \,\textrm{d}\overline{\Omega},
\label{elEnergy}
\end{equation}
where $\text{d}\overline{\Omega}=\|\overline{\boldsymbol{a}}_1 \times \overline{\boldsymbol{a}}_2 \| \textrm{d}\theta^1 \textrm{d}\theta^2$. The elastic tensor is given component-wise by
\begin{equation}
C^{i j k l}=\frac{E_\text{s}}{1-\nu_\text{s}^2}\left(\nu_\text{s} \overline{a}^{i j} \overline{a}^{k l}+
\frac{1-\nu_\text{s}}{2} \left(\overline{a}^{i k}\overline{a}^{j l}+\overline{a}^{i l}\overline{a}^{j k} \right)\right)
\end{equation}
in curvilinear coordinates. Inertial forces are included by adding a kinetic energy contribution of the form 
\begin{equation}
U_k \left[\boldsymbol{x}\right]=
\frac{1}{2}\int_{\overline{\Omega}}h_\text{s} \rho_\text{s} \dot{\boldsymbol{x}} \cdot \dot{\boldsymbol{x}} \,\text{d}\overline{\Omega},
\label{kinEnergy}
\end{equation}where $\rho_\mathrm{s}$ is the shell mass density and $\dot{\boldsymbol{x}} =\partial \boldsymbol{x} / \partial t$ is the velocity.

The total energy $U=U_e+U_k$ is minimized with the FEM in weak formulation. To allow for a mathematically sound representation of the bending field, the shape functions need to be differentiable with continuous derivatives (class $C^1$) over the whole domain. We satisfy this requirement by adopting the subdivision surface paradigm~\cite{Cirak2000,Cirak2001}. The use of subdivision surface shape functions avoids the introduction of auxiliary degrees of freedom such as rotations and is therefore particularly efficient. The dynamic problem is solved through time integration with a predictor-corrector scheme from the Newmark family \cite{Newmark1959}.

\section{Coupling schemes for the hybrid approach}

\subsection{Debris flow as hybrid media: the LBM-DEM coupling} \label{LBM-DEM}
Grains interact with the fluid through a forcing term acting on the LBM velocity field. To achieve this, we employ a simplified version of the Immersed Boundary Method \cite{Owen2011} that has the basic assumption that the fluid fills the entire domain, including the interior of the grains. The volumetric displacement of the fluid due to the presence of the grains is therefore neglected. Fluid elements located away from the grains are not directly influenced by the coupling. All nodes located inside of a grain, on the other hand, are subjected to a forcing term $\boldsymbol{p}$ that relaxes the velocity of the fluid to the velocity of the grain. This is included together with gravity $\boldsymbol{g}$ in Eq.~\ref{newfunctions} inside the forcing term  $\boldsymbol{F}=\boldsymbol{g}+\boldsymbol{p}$. For all nodes located outside of grains, this reduces to  $\boldsymbol{F}=\boldsymbol{g}$. If $\boldsymbol{x}$ is the location of the fluid node, $\boldsymbol{p}$ can be computed, assuming the unit volume of a lattice node, as
\begin{equation}
 \boldsymbol{p}(\boldsymbol{x},t)={\rho_\text{f}(\boldsymbol{x},t)}\left[ \boldsymbol{u}_\text{f}(\boldsymbol{x},t) - \boldsymbol{u}_\text{g}(\boldsymbol{x},t)\right].
\end{equation}
$\boldsymbol{u}_\text{f}(\boldsymbol{x},t)$ denotes the velocity of the fluid, and $\boldsymbol{u}_\text{g}(\boldsymbol{x},t)$ is the velocity of the grain at the same position. Since grains are rigid bodies, $\boldsymbol{u}_\text{g}(\boldsymbol{x},t)$ can be calculated as
\begin{equation}
 \boldsymbol{u}_\text{g}(\boldsymbol{x},t)=\boldsymbol{u}_\text{g}(\boldsymbol{x_\text{c}},t) +(\boldsymbol{x}-\boldsymbol{x}_c) \times \boldsymbol{\omega}_g,
\end{equation}
where $\boldsymbol{v}_g$ and $\boldsymbol{\omega}_g$ are the translational and rotational velocities of the grain respectively, and $\boldsymbol{x}_c$ is the position of its center of mass. To improve the stability of this scheme, the viscosity of the fluid lying inside the grain is set to a high value.

The forcing term $\boldsymbol{F}$ is solved, following the approach of Ref.~\cite{Guo2002}, through the addition of the extra term $F_i (\boldsymbol{x},t,\boldsymbol{F})$ in Eq.~\ref{newfunctions}, and modifying the computation of the velocity field in Eq.~\ref{densityVelocity} with
\begin{equation}
 \boldsymbol{u}_\text{f}=\frac{1}{\rho_\text{f}}\left(\sum\limits_i f_i\boldsymbol{c}_i  +\frac{\boldsymbol{F}}{2}\right).
 \end{equation}
The same force with an opposite sign is applied to the grains. The overall force $\boldsymbol{F}_\text{grain}$ and torque $\boldsymbol{T}_\text{grain}$ acting on an element is therefore the sum of all contributions from the nodes $j$ lying inside the grain itself:
\begin{equation}
\boldsymbol{F}_\text{grain}= -\sum_j{ \boldsymbol{p}_j(\boldsymbol{x}_j)},\qquad
\boldsymbol{T}_\text{grain}= -\sum_j{ \boldsymbol{p}_j(\boldsymbol{x}_j)} \times (\boldsymbol{x}_j-\boldsymbol{x}_c).
\end{equation}
This is transmitted to the DEM solver and adds up to the collection of forces for the equations of motion.

\subsection{The shell as flexible barrier: the DEM-FEM coupling} \label{DEM-FEM}
When a grain collides with the shell, a repulsive force is exerted on both elements, following an approach similar to the grain-to-grain contact of Sec.~\ref{DEM}. For this purpose, a set of points are generated from the deformed state of the shell $\bar{\boldsymbol{x}}$, and are used for the resolution of contacts. This approach mimics the actual behavior of a cable-net barrier, since grains smaller than the point spacing are allowed to pass through.

The overlap between a grain $g$ of radius $r_\text{g}$ and a shell point $p$ is calculated in a similar way to the grain-grain overlap:
\begin{equation}
 \xi=r-\left\|\boldsymbol{d}_{\text{p},\text{g}}\right\|,
\end{equation}
where $\boldsymbol{d}_{\text{p},\text{g}}$ denotes the distance vector between the center of the grain and the shell point. When positive, this overlap is used to calculate a normal repulsion force $F^\text{n}_{\text{obstacle}}$ using Eq.~\ref{hertz}.

A static friction component is added through a spring introduced in the plane ortogonal to $\boldsymbol{d}_{\text{p},\text{g}}$, in order to model the trapping effect of the barrier. The spring is initialized at the time of initial contact with the barrier $t_\text{init}$ and is removed when a limit elongation is reached. The elongation
\begin{equation}
\zeta=\int^t_{t_\text{init}} u_{\text{rel}}^\text{t} \,\textrm{d}t,
\end{equation}
is used to determine with the spring stiffness $k$ the restoring force as
\begin{equation}
 F^\text{t}_{\text{obstacle}}=-\textrm{sign}\left(u_{\text{rel}}^\text{t}\right)
 \min\left\{k \vert \zeta \vert, \tan (\psi) F^\text{n}_{\text{obstacle}}\right\}.
\end{equation}
The resulting force is transmitted to both the colliding DEM grain and the FEM mesh.

We chose contact points to be coincident to mesh nodes. This avoids interpolations between the contact points and the mesh nodes. Since we use a regular mesh, the characteristic filtering properties of the barrier are determined by the element size. Note that this makes the contact physics dependent on the mesh resolution, which could be considered undesirable. However, in our simulations, the overall mesh size, as determined by a convergence study, is already rather fine. Our approach is motivated by simplicity and could well be refined without changes to the methodology. If higher precision is needed, the limit surface of the subdivision shell instead of the deformed control mesh can be used. This comes at the cost of longer simulation times, but allows for a higher precision~\cite{Vetter2013}.

\section{Reference geometry of a flexible barrier model}
\label{setting}

\begin{figure}[tb]
\centering
\includegraphics[]{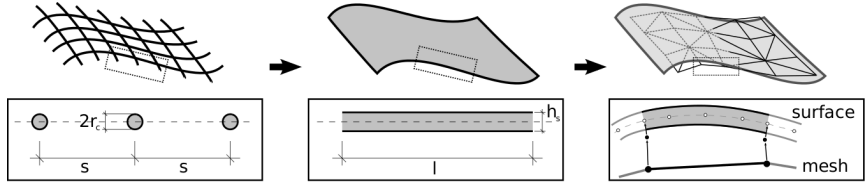}
\caption{\label{eqMembrane} The cable-net structure of the barrier is modeled with the FEM as a shell with equivalent stiffness, see Eq.~\ref{membraneStiffness}.}
\end{figure}

\begin{figure}[tb]
\centering
\includegraphics[width=0.5\textwidth]{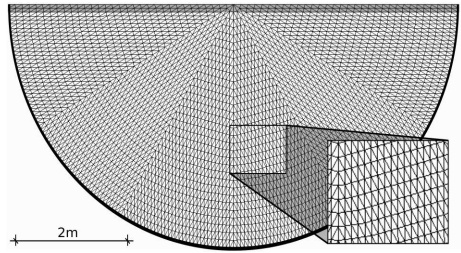}
\caption{\label{shellMesh}Discretization of the shell mesh. The half circle is pinned along the solid black line (translation is blocked, rotation is allowed) and has a free edge at the top rim. The darker layer at the free edge has a higher stiffness.}
\end{figure}

Our simulations are carried out following the geometries of in-situ experiments, of the sort of Ref.~\cite{Bugnion2011a}. We reproduce a gully riverbed of cylindrical shape with a diameter of $W=8\ \text{m}$, see Fig. \ref{show}. The debris center of mass $C$ is positioned at a variable distance $R$ from the barrier and is instantaneously released at the beginning of the simulation, similarly to the procedure recommended for the dam-break test. The front of the flow is artificially enriched with grains to resemble the observed impact conditions. The material then accelerates under the effect of gravity, which has a component both in the $x$ direction (the longitudinal direction of the channel) and in the $z$ direction (vertical). The ratio between the two accelerations gives the inclination of the channel, which is fixed to $15^\circ$.  The channel is $22\ \text{m}+R$ long and is loaded with $140\ \text{m}^3$ of fluid. The chosen non-Newtonian rheological law for the fluid is the Bingham plastic, which is the most commonly adopted when describing mudflow rheology \cite{Kaitna2007} and can be easily implemented with the LBM \cite{Leonardi2014a}. It is defined by
\begin{equation}
   \left\{
  \begin{array}{l l}
    \dot{\boldsymbol{\gamma}}=0 & \quad \textrm{if fluid has not yielded,}\ (\left\|\boldsymbol{\sigma}\right\|<\sigma_{\mathrm{y}})\\
    \boldsymbol{\sigma} = \sigma_{\mathrm{y}} \frac{ \dot{\boldsymbol{\gamma}}}{ \left\|\dot{\boldsymbol{\gamma}}\right\|} +2\mu_\mathrm{pl} \dot{\boldsymbol{\gamma}} & \quad \textrm{if fluid has yielded}\ (\left\|\boldsymbol{\sigma}\right\|>\sigma_{\mathrm{y}}),\\
  \end{array} \right.\
\end{equation}
where $\dot{\boldsymbol{\gamma}}$ and $ \boldsymbol{\sigma}$ are the shear rate and the shear stress tensor, respectively, and the vertical bars $\left\|\cdot\right\|$ denote their magnitudes (i.e.~the second invariant). The scalars $\sigma_{\mathrm{y}}$ and $\mu_\mathrm{pl}$ are the yield stress and the plastic viscosity. All results shown in the next section are obtained with $\sigma_{\mathrm{y}}= 500\ \text{Pa}$, $\mu_\mathrm{pl}=50\ \text{Pa}/\text{s}$ and fluid mass density $\rho_\text{f}=1500 \ \text{kg}/\text{m}^3$. The fluid phase is mixed with a variable amount of grains up to $34\ \text{m}^3$. The grain radii are sampled randomly from a uniform distribution between $0.1$ and $0.25~\textrm{m}$ with a mass density of $\rho_\text{g}$ = 3500~kg/m$^3$.

The actual barrier is usually a complex reticular structure, made of diverse combinations of steel cable-nets of different sizes and shapes. In this work, we consider a simplified structure, as shown in Fig.~\ref{eqMembrane}, composed of steel cables with radius $r_\text{c}$, Young's modulus $E_\text{c}$, and regular spacing $s$. The barrier is modeled using shell elements, whose material and geometric characteristics need to be determined to match those of a cable net. We do so by setting the Young's modulus and the thickness of the shell to equivalent values $E_\text{s}$ and $h_\text{s}$. These values are calculated by imposing that a structural element of unit length ($l=1$) have the same stretching stiffness as the equivalent net. We obtain: 
\begin{equation}
 \pi E_\text{c} r_\text{c}^2 \frac{l}{s}=h_\text{s} E_\text{s} l.
 \label{membraneStiffness}
 \end{equation}
The simulations of the next chapter aim to reproduce a barrier with $s=0.3 \ \text{m}$, $r_\text{c}=1.1 \ \text{mm}$, and $E_\text{c}=200 \ \text{GPa} $. This is achieved by imposing $E_\text{s}=0.26 \ \text{GPa}$ and $ h_\text{s}=0.01 \ \text{m}$.

Actual debris flow barriers are complemented with reinforcement cables of different sizes in order to increase the stiffness at critical points. This is particularly important at the upper rim of the barrier to avoid excessive overspill and to improve the retention. To reproduce the same behavior without implementing further models, the upper elements of the shell are stiffened up to 10 times the value of the rest of the barrier (see Fig.~\ref{shellMesh}). Note that linear, isotropic behavior is only a crude approximation of the reticular mechanical behavior. It can be replaced by the characteristic non-linear, anisotropic, plastic behavior of specific wire mesh configurations and orientations under large deformations at any time.

\section{Results of debris impacts on the barrier} \label{results}

\begin{figure}[!ht]
\centering
\includegraphics[]{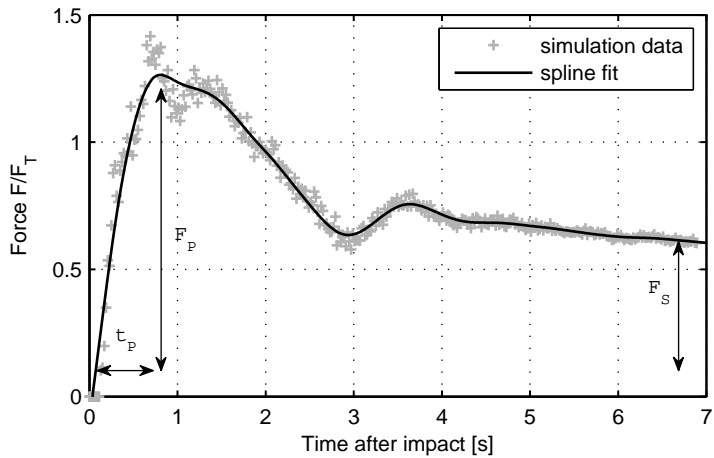}
\caption{\label{evolution} Typical force evolution, depicting the total action of the debris on the net in the longitudinal direction of the channel.}
\end{figure}

The principal value of interest is the force $F$ transmitted from the flow to the barrier and specifically the component ortogonal to the deformed surface (see Fig.~\ref{evolution}). It is calculated by a sum of the FEM-DEM contact forces over all grains. The force evolution over time always presents an initial sharp increase followed by the relaxation to a stationary value.
We analyze the force evolution by fitting a spline to the numerical data, which is then differentiated to obtain the peak force $F_\textrm{P}$ (corresponding to the first inflection point), the delay of the peak from the moment of initial impact $t_\text{S}$, and the stationary force at the tail $F_\textrm{S}$. The observed evolution is consistent with earlier experimental findings \cite{Moriguchi2009,Chanut2010,Teufelsbauer2011,Faug2011,Caccamo2012}. Due to the flexible nature of the barrier the burst duration $t_\text{H}$ is large compared to rigid obstacle impact. All force measures are presented in a dimensionless form, through division by the static load $F_\textrm{T}$ in the $x$ direction, calculated as
\begin{equation}
F_\text{T}=m_\text{T} g \sin(15^\circ),
\end{equation}
where $m_\text{T}$ denotes the total mass of the debris (grains and fluid) and $g$ the gravity.

\begin{figure}[!ht]
\centering
\includegraphics[]{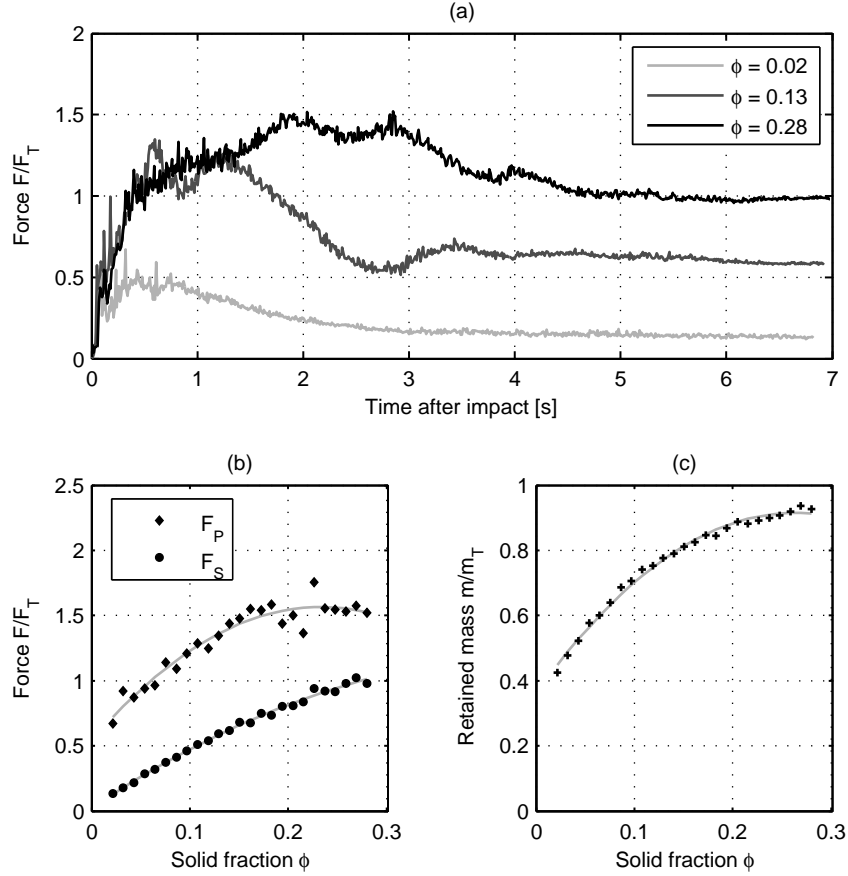}
\caption{\label{solidFractionPlot} Characterization of the impact force evolutions as function of the grain content $\phi$ (a). Stationary and peak force increase as the grain content becomes higher (b). Also the retained mass increases, approaching nearly total retainment for higher grain contents (c).}
\end{figure}

We study the barrier defined in the previous section. In the first simulation set, the force evolution is obtained using a variable number of grains, i.e., by varying the grain content
\begin{equation}
\phi=V_{\text{grains}}/\left(V_{\text{grains}}+V_{\text{fluid}}\right).
\end{equation}
The number of grains $N_{\text{grains}}$ is between $100 - 1300$, and therefore $\phi$ is between $0.02 - 0.28$. Fig.~\ref{solidFractionPlot} (a) shows how the force evolution changes with the increase in grain content. The higher the grain content, the quicker the barrier permeability is reduced. For this reason, the peak value $F_\text{P}$ in Fig.~\ref{solidFractionPlot} (b) is increasing with $\phi$. Furthermore, a quicker reduction of the barrier permeability means that the amount of material impounded by the barrier itself is increasing, leading to a growing stationary force $F_\textrm{S}$, too.

\begin{figure}[!ht]
\centering
\includegraphics[]{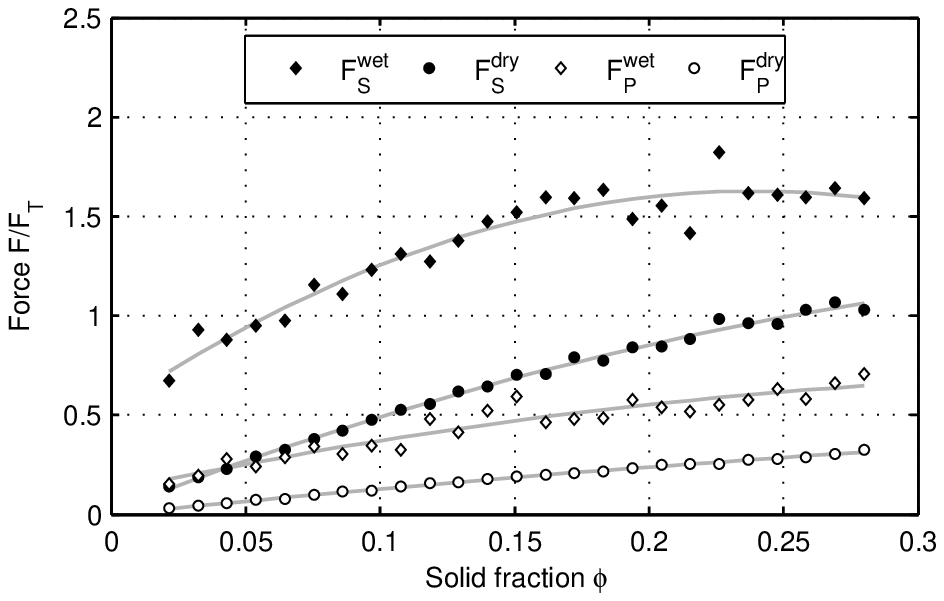}
\caption{\label{dryWetPlot} Comparison between simulations with fluid (diamonds) and without fluid (circles).}
\end{figure}

In order to understand the influence of the fluid phase on the force evolution, we perform two series of simulations: wet (with fluid), and dry (only DEM, no fluid). The difference between the wet and the dry force evolutions with identical grain content is shown Fig.~\ref{dryWetPlot}, exhibiting the resulting difference in peak and stationary forces. In the wet case, forces are much higher, which is surprising for a model that does not explicitly take the interaction between fluid and barrier into account. The increase in peak values is induced by the hydrodynamic interaction between grains and fluid, which in turn transforms into a higher momentum transfer between grains and barrier. The increase in stationary values is induced by the fluid that is retained by the barrier after the barrier itself is saturated with grains. A higher grain content means a quicker saturation of the barrier and an increase in the amount of retained fluid. The bottom right panel of Fig.~\ref{solidFractionPlot} shows how the retained mass of material (both fluid and grains) increases with higher grain contents. The combination of the results in Figs.~\ref{solidFractionPlot} and~\ref{dryWetPlot} suggests that the design of a barrier should always be tested against a fluid debris flow with a high grain content. This challenges the design procedures based on the simplification of modeling the debris flow as a purely discrete or purely liquid material. We show how none of the phases can be neglected, when calculating the impact force. Because of this observation, all results shown in the following are obtained using the hybrid model with the highest particle content ($\phi=0.28$).

\begin{figure}[!ht]
\centering
\includegraphics[]{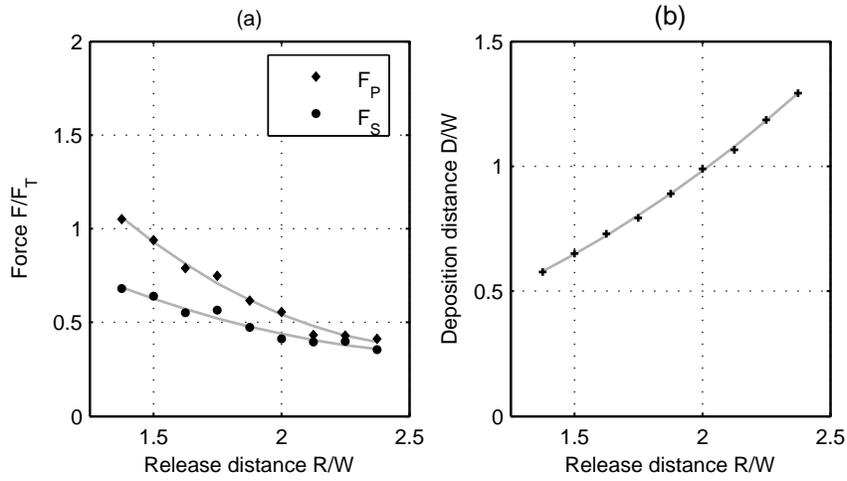}
\caption{\label{distancePlotSmall} Characterization of the force evolution as function of the release distance from the barrier $R$. The deposition of material in the channel lowers the stationary and peak force (a). The final deposition distance $D$ is shown on panel (b).}
\end{figure}

The motion of the debris mass can be conceptually divided into two parts. In the first, the material accelerates, while in the second part the flow velocity is reduced and the deposition process takes place. To study the effect of the deposition on the force evolution we perform a set of simulations, each differing in the release distance $R$, between $10 -20~\textrm{m}$. The results (Fig.~\ref{distancePlotSmall}) show how the initial peak, which can be assumed to be proportional to the dynamic load, is steeply decreasing for larger distances. The stationary load is also decreasing, indicating that more and more material is deposited in the channel before the barrier. Fig.~\ref{distancePlotSmall}(b) shows the final distance of the debris center of mass from the barrier. The further away from the barrier the debris is released, the more of it deposits in the channel before the barrier, leading to a lower force impact.
 
\begin{figure}[!ht]
\centering
\includegraphics[]{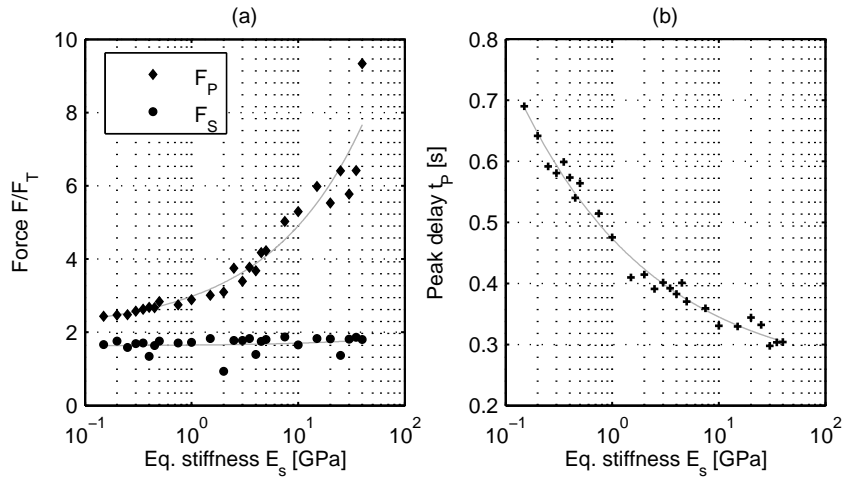}
\caption{\label{newStiffness} Characterization of the force evolution as function of the barrier stiffness. Note how the peak force, related to the dynamic load, decreases for a flexible structure (a). A stiffer structure is subjected to a stronger and quicker initial burst, as can be seen from panel (b). }
\end{figure}

Traditional barriers are modeled as rigid obstacles, enabling the designer to consider the structural and the hydrodynamic problems separately. Most design guidelines are based on an impact force estimation that relies on this hypothesis \cite{Hubl2009}. The maximum impact force is generally calculated as a function of the dynamic load, proportional to the square of the flow velocity. In flexible barriers, this dynamic load can be consistently reduced by the flexibility of the structure. To show this, we present a set of simulations sharing the same debris configuration, and therefore the external action, but differing in the stiffness of the barrier. We vary the stiffness by three orders of magnitude, from a minimum shell Young's modulus $E_\text{s}$ of $0.05 \ \text{GPa}$ to a maximum of  $50 \ \text{GPa}$. (see Fig.~\ref{newStiffness}). The same debris flow transmits a much higher peak force to a stiffer structure, while the stationary force shows no dependence. The amount of impounded material is the same, but a flexible structure is safer from structural collapse. The mechanism of force reduction is revealed by the right panel of  Fig.~\ref{newStiffness}. The duration of the initial burst is longer for more flexible structures, allowing for an adsorption of the dynamic load over longer times, and therefore reducing its peak.

 \begin{figure}[!ht]
	\begin{subfigure}{\textwidth}
        \centering
        \includegraphics[]{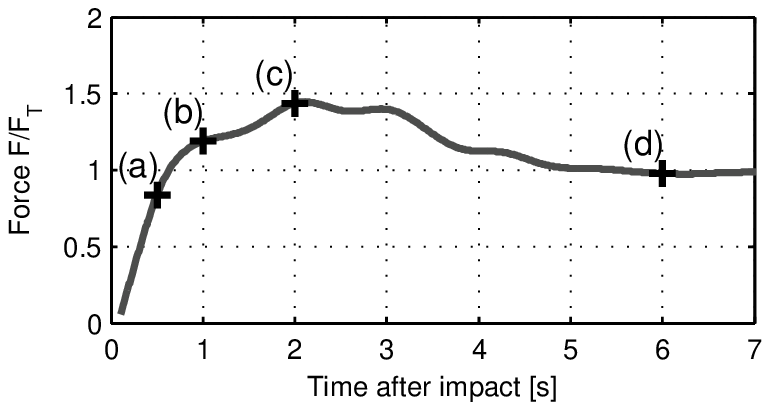}
    \end{subfigure}
    \begin{subfigure}{\textwidth}
        \centering
        \includegraphics[]{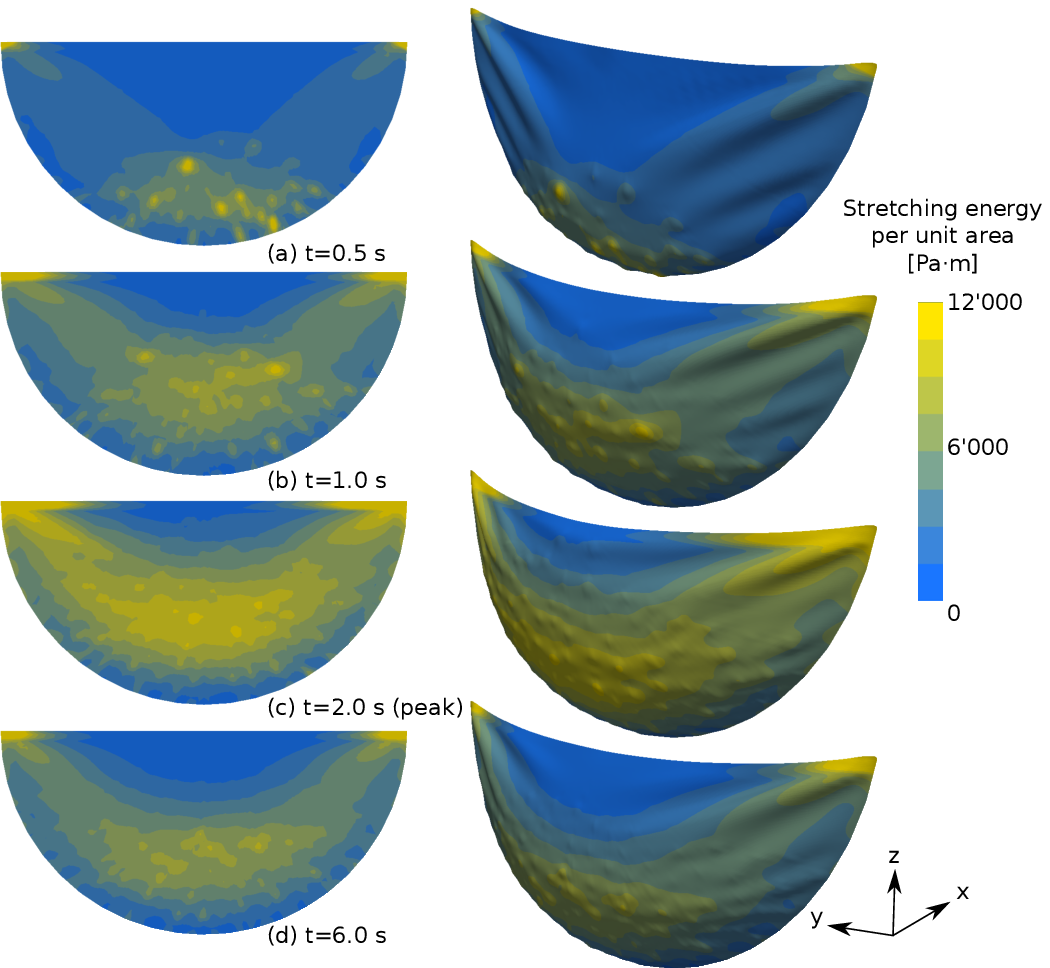}
    \end{subfigure}
    \caption{\label{barrier} Streching energy distribution in the barrier, for $\phi=0.28$ and $R=10\ \text{m}$. The considered time steps are indicated by the orange points in the force evolution. The pictures on the right are renderings of the deformed shell, while the pictures on the left have no deformation nor 3D shadings, for clarity.}
\end{figure}

A key feature of our numerical framework is the capability to yield information about the stress distribution in the barrier. The deformed configuration of the barrier, according to the simulation with $\phi = 0.28$ and $R=10\ \text{m}$ is shown in Fig.~\ref{barrier}. The color contours indicate the elastic energy per unit area of the shell, as obtained from Eq.~\ref{elEnergy}. It clearly shows how the stress is localized at the upper rim of the barrier, where the shell has been stiffened. The force distribution on the supports can be inferred from this, showing how the upper supports are the ones under the highest and possibly critical load.

\section{Summary and Outlook}
A computational framework has been established, able to couple the FEM representation of a cable-net barrier with an idealized debris flow. The debris flow is obtained through a further coupling between the DEM and the LBM, with both the granular component and the fluid explicitly represented.
The missing coupling between the fluid (LBM) and the barrier (FEM), which should be the least significant interaction due to the cable net permeability, has been neglected. Nevertheless, these two components indirectly interact through mutual coupling with the granular phase (DEM), which transmits hydrodynamic forces to the barrier. The results show how both granular and fluid component of the debris flow have a key impact on the force evolution, challenging design assumptions based on neglecting one of the two phases. The effect of the flexibility of the structure has also been studied, showing how a flexible barrier is more efficient in reducing the peak impact force, and in distributing the dynamic load over a longer time. Further work in this direction will focus on designing the granular phase based on field data and understanding how to calibrate the filtering properties of the barrier based on the grain characteristics and vice versa. For the same reason, the filtering properties of the barrier should be independent of the shell discretization, in order to leave the freedom to choose the mesh size only based on convergence criteria. Currently we are adding material anisotropy and plasticity to the shell approach, that along with non-linear elasticity, e.g. a Green elastic material behavior, are the main elements for representing homogenized wire mesh mechanics.

\section*{Acknowledgement} The research leading to these results has received funding from the European research network MUMOLADE (Multiscale Modelling of Landslides and Debris Flow), the ETH Zurich by ETHIIRA grant no.~ETH-03 10-3, as well as from the European Research Council Advanced Grant no.~319968-FlowCCS. We thank Corinna Wendeler from Geobrugg AG - Geohazard Solutions for valuable discussions.

\clearpage
\begin{appendix}
\bibliographystyle{ieeetr}
\bibliography{mendeley}
\end{appendix}

\end{document}